**Unique magnetic structure of the vdW antiferromagnet VBr$_3$**

Milan Klicpera[1,*], Ondřej Michal[1], Dávid Hovančík[1], Karel Carva[1], Oscar Ramon Fabelo Rosa[2], M. Orlita[3], Vladimír Sechovský[1], and Jiří Pospíšil[1]

[1]*Charles University, Faculty of Mathematics and Physics, Department of Condensed Matter Physics,*
*Ke Karlovu 5, 121 16 Prague 2, Czech Republic*
[2]*Institut Laue-Langevin, 71 avenue des Martyrs, CS 20156, 38042 Grenoble Cedex 9, France*
[3]*LNCMI, UPR 3228, CNRS, EMFL, Université Grenoble Alpes, 38000 Grenoble, France*

[*]Corresponding author's email address: *milan.klicpera@matfyz.cuni.cz*

ABSTRACT

VBr$_3$ is a van der Waals antiferromagnet below the Néel temperature of 26.5 K with a saturation moment of 1.2 µ$_B$/f.u. above the metamagnetic transitions detected in the in-plane and out-of-plane directions. To reveal the AFM structure of VBr$_3$ experimentally, we performed a single-crystal neutron diffraction study on a large high-quality crystal. The collected data confirmed a slight monoclinic distortion of the high-temperature rhombohedral structure below 90 K. The magnetic structure was, nevertheless, investigated within the *R-3* model. The antiferromagnetic structure propagation vector $k = (1, 0, ½)$ was revealed. In an attempt to determine the magnetic structure, 72 non-equivalent magnetic reflections were recorded. The experimental data were confronted with the magnetic space groups dictated by the *R-3* lattice symmetry and propagation vector. The best agreement between the experimental data and the magnetic structure model was obtained for the space group *P-1.1'_c*. The magnetic unit cell of the proposed unique antiferromagnetic structure with periodicity 6c is built from two identical triple layers antiferromagnetically coupled along the c axis. Each triple layer comprises a Néel antiferromagnetic monolayer sandwiched between two antiferromagnetically coupled ferromagnetic monolayers.

## 1. Introduction

Magnetism in two dimensions is one of the most interesting phenomena in condensed matter physics. Atomically thin 2D materials provide a promising platform for investigating magnetic properties applicable in spintronic devices, spin-dependent optoelectronics, and photonics [1-3]. Recent research into 2D antiferromagnets (AFMs) has led to significant advances in applications based on materials with zero net magnetic moment but are intrinsically magnetic. Two-dimensional van der Waals (vdW) materials with magnetic ions on a honeycomb lattice exhibit various exotic behaviours.

Anisotropic exchange interactions play a key role in creating the magnetic structure of layered 2D antiferromagnets [4]. Depending on the hierarchy of interlayer and intralayer exchange interactions, different AFMs can be divided into two categories: those with intralayer AFM order and those with interlayer AFM order [5, 6]. Those with intralayer AFM coupling are allowed to have the interlayer coupling either ferromagnetic (C-type AFMs) or antiferromagnetic (G-type AFMs). Intralayer AFMs can be further distinguished as AFM-Néel, AFM-zigzag, and AFM-stripy states [7, 8]. The A-type AFMs consist of AFM-coupled ferromagnetic layers [9].

An example, $CrPS_4$, has an A-type AFM ground state formed by out-of-plane polarized ferromagnetic monolayers AFM coupled along the c-axis. Due to small magnetic anisotropy energy and weak interlayer coupling, a spin-flop metamagnetic transition (MT) appears at 0.7 T for the field parallel to the c-axis, which is followed by a spin-flip transition at 8 T to a polarized paramagnet [10].

Several intriguing intralayer AFMs have been revealed among $MPS_3$ ($M$ = Mn, Fe, Co, or Ni) and $MPSe_3$ ($M$ = Mn, Fe, or Ni) compounds [11, 12]. $FePS_3$, $CoPS_3$, $NiPS_3$, and $FePSe_3$ display zig-zag-type AFM ordering [13-17], whereas Néel-type ordering is seen in $MnPS_3$ and $MnPSe_3$ [11, 12, 15]. $FePS_3$, $FePSe_3$, and $MnPS_3$ have out-of-plane anisotropy [14, 16, 18, 19] while $NiPS_3$ and $MnPSe_3$ are characterized by in-plane anisotropy [20, 21].

Our research activities are focused on the magnetism in the transition metal trihalides $CrX_3$ and $VX_3$ ($X$ = I, Br, Cl) that exhibit diverse structural and magnetic properties. All of them are polymorphic [22, 23]. The $CrX_3$ α-phases crystallize in the $BiI_3$ type trigonal structure (space group *R-3*), and the β-phases are monoclinic of the $AlCl_3$ type (*C2/m*) at temperatures above 200 K. On the contrary, the high-temperature phases of $VX_3$ analogues are the $BiI_3$-type trigonal and undergo a transition to a monoclinic structure when cooling below 100 K [23-25].

$CrCl_3$ [26-29], $VBr_3$ [30, 31] and $VCl_3$ [23, 32] are antiferromagnetic, while the other three order ferromagnetically [25, 33, 34]. The compounds of the two Cr and V groups of trihalides exhibit different characteristic moments of trivalent transition metal ions. $Cr^{3+}$ ions possess a quenched orbital moment ($S = 3/2$, $L = 0$), leading to the ordered moment of 3 $\mu_B$ per ion [33, 35-37]. The significantly reduced saturated magnetic moment of 1.2 $\mu_B$ per vanadium ion observed for $VI_3$ differs from the pure spin state ($S = 1$, $L = 0$; $\mu_{sat} = 2$ $\mu_B/V^{3+}$). This result can be understood by considering an unquenched orbital moment as proposed by the theory [38] and subsequently confirmed by our XMCD measurements [39].

Regarding antiferromagnets, $CrCl_3$ exhibits A-type AFM with in-plane polarized magnetic moments [26-28]. $VBr_3$ is AFM below $T_N = 26.5$ K [30, 31]. The temperature dependence of the magnetic susceptibility measured at low magnetic fields is typical for AFM. Still, it shows an as yet unexplained anomalous increase with cooling below 15 K. Magnetization data collected at low temperatures revealed a sharp metamagnetic transition in a considerably high magnetic field of 16.9 T in out-of-plane orientation. In contrast, the response to the in-plane field exhibits a continuous tilting of the rotation, which is completed in the anisotropy field 27 T [30]. Such high values of the critical fields are typical for intralayer coupled AFMs [21, 40] rather than for A-type AFMs. Both magnetization curves reach the same value of saturated magnetic moment $\mu_{sat} \approx 1.2$ $\mu_B$/f.u. [30] that compares to the value revealed in the case of $VI_3$ [25]. These results signalize both the presence of an unquenched orbital moment and an intralayer AFM structure. The AFM zig-zag magnetic structure was recently proposed by theory, based on magnetization data, as a favourable one considering the Heisenberg model with interactions up to the third nearest neighbour ($J_1$-$J_2$-$J_3$ model) [30].

Knowledge of the magnetic structure is essential for better understanding the physics of vdW AFM VBr$_3$. Therefore, we have performed a neutron diffraction experiment on a high-quality VBr$_3$ single crystal. Based on a detailed analysis of the collected neutron diffraction data, a unique compensated long-period antiferromagnetic structure combining Néel AFM monolayers with ferromagnetic monolayers was proposed and discussed in the present study.

## 2. Experimental

The single crystals of VBr$_3$ have been grown by the chemical vapor method from pure elements, as detailed in our previous work [30]. This synthesis procedure prevented any contamination of the final product by residuals from precursors readily used to produce Br$_2$, e.g. TeBr$_4$ [41] or VBr$_2$O[31] employed in the reaction and transporting tube. The plate-like single crystals with an edge of more than 10 mm long were grown from the pre-reacted mixture of the stoichiometric amount of pure elements (V 99.9 %, Br$_2$ 99.5%) by a slow recrystallization process [42]. The sample quality, orientation and structure of single crystals were verified by the Laue method; see Figure S1 in Supplementary materials. The low-temperature properties were studied by specific heat measurements, which identified the temperatures of structural (90 K) and AFM transition (26.5 K) in perfect agreement with previously published results [23, 30, 31].

A neutron diffraction experiment was carried out on a plate-like single crystal with dimensions of approximately 10.0 x 3.0 x 0.1 mm$^3$ using the D10 four-circle neutron diffractometer at the Institute Laue-Langevin (ILL), Grenoble. A pyrolytic graphite monochromator selecting the neutron wavelength of $\lambda$ = 2.36 Å and a 94 x 94 mm$^2$ area detector with spatial resolution 1.5 x 1.5 mm$^2$ were employed in the experiment.

To prevent any degradation of the moisture-sensitive sample [43] during its transport to the instrument, it was sealed under an Ar atmosphere. The sample was removed from the protective container and inserted into the He flushed cryostat at D10, minimizing the time it spent on air to a few seconds. After the experiment, it was removed from the cryostat and inserted into the container under a continuous flow of the He gas. Subsequently, the sample was sealed under an Ar protective atmosphere. No sample degradation was observed, allowing further investigations.

The neutron diffraction measurements began with determining the UB matrix at room temperature based on a set of 23 nuclear reflections of the respective *R-3* lattice in the hexagonal description. The NOMAD software was used for instrument control and data collection. The sample was cooled down to 100 K while the intensity of reflections (1, 1, 0), (3, 0, 0), and (0, 0, 6) was followed. The UB matrix was optimized based on these reflections, and the temperature dependence of their intensity was measured on further cooling. Finally, the UB matrix was determined at 2 K based on the original 23 reflections, still considering the hexagonal unit cell corresponding to the *R-3* space group. Only a slight change of matrix parameters was observed on cooling, being consistent with a standard thermal contraction.

The magnetic signal was searched for by performing q-scans along principal crystallographic directions of the hexagonal lattice at base temperature. Subsequently, both nuclear and magnetic reflections were measured using standard ω-scans. The collected data were treated using the FullProf suite [44]. Irreducible representations and magnetic space groups of the paramagnetic *R-3* group were calculated employing the Isodistort suite [45].

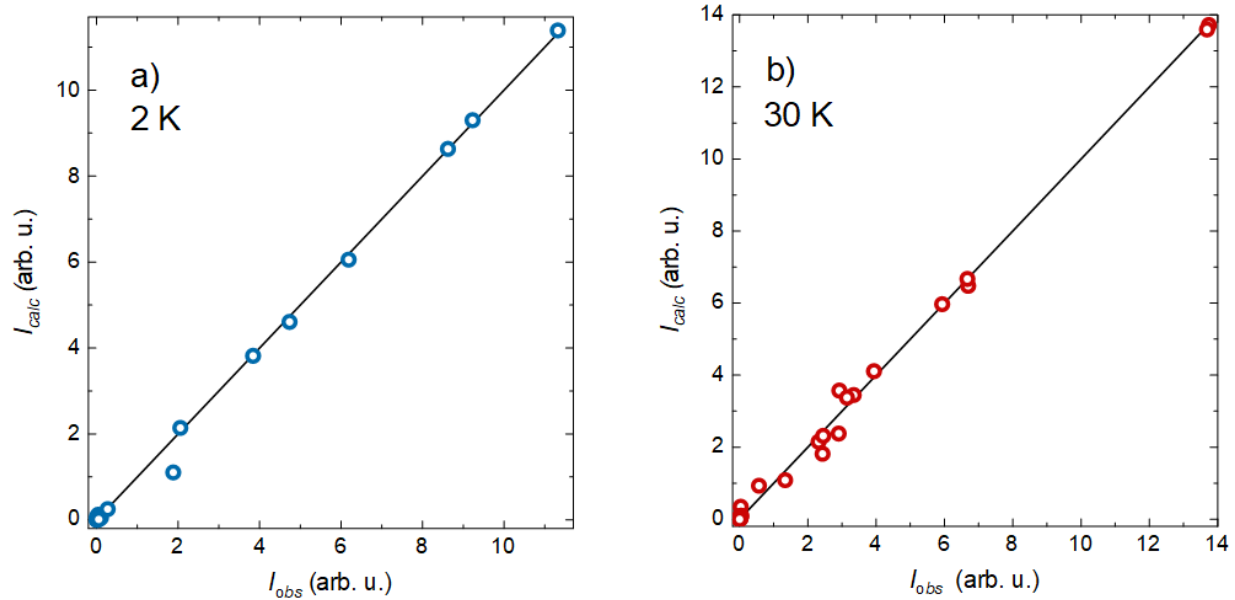

Figure 1 – Intensity on nuclear diffractions measured at a) 2 K and b) 30 K using the D10 instrument at the ILL Grenoble. A comparison of measured intensity and intensity calculated employing the model *R-3* structure is presented. Only scale, extinction parameters and thermal factors were fitted, leading to agreement factors $R_F$ of 3.4 % and 6.0 %, respectively.

## 3. Results

$VBr_3$ single crystal was verified to crystallize in the rhombohedral *R-3* crystal structure at room temperature using the volume neutron diffraction technique. The determined lattice parameters in the hexagonal notation $a = 6.3824(5)$ Å and $c = 18.4384(5)$ Å are in agreement with previous X-ray diffraction studies [23, 31]. The high-temperature rhombohedral phase of $VBr_3$ has been shown to undergo a structural phase transition to a, so far unresolved, lower symmetry structure below 90 K [23, 24, 31]. Indeed, following the nuclear reflections (1, 1, 0), (3, 0, 0), and (0, 0, 6) of the *R-3* structure at temperature revealed a small broadening of respective diffraction peaks below 90 K, evidencing a slight distortion during the structural phase transition; see Figure S2 in Supplementary materials [46]. Additional nuclear reflections outside *R-3* positions were observed in lower-temperature neutron data. A triclinic lattice with hexagonal-like parameters was considered in searching for lower-symmetry nuclear reflections above the magnetic ordering temperature of 26.5 K [30, 31], namely at 30 K. Only very few *P1* reflections were found. Moreover, their intensity was significantly reduced compared to the strong *R-3* reflections. Determination of the crystal structure of the low-temperature phase, therefore proved to be infeasible. The excellent agreement (Figure 1) between the measured data, at both 30 K (146 nuclear reflections, 39 inequivalent reflections) and 2 K (36 nuclear reflections, 17 inequivalent reflections), and the rhombohedral structure model demonstrates that also the low-temperature phase can be effectively described by the *R-3* space group.

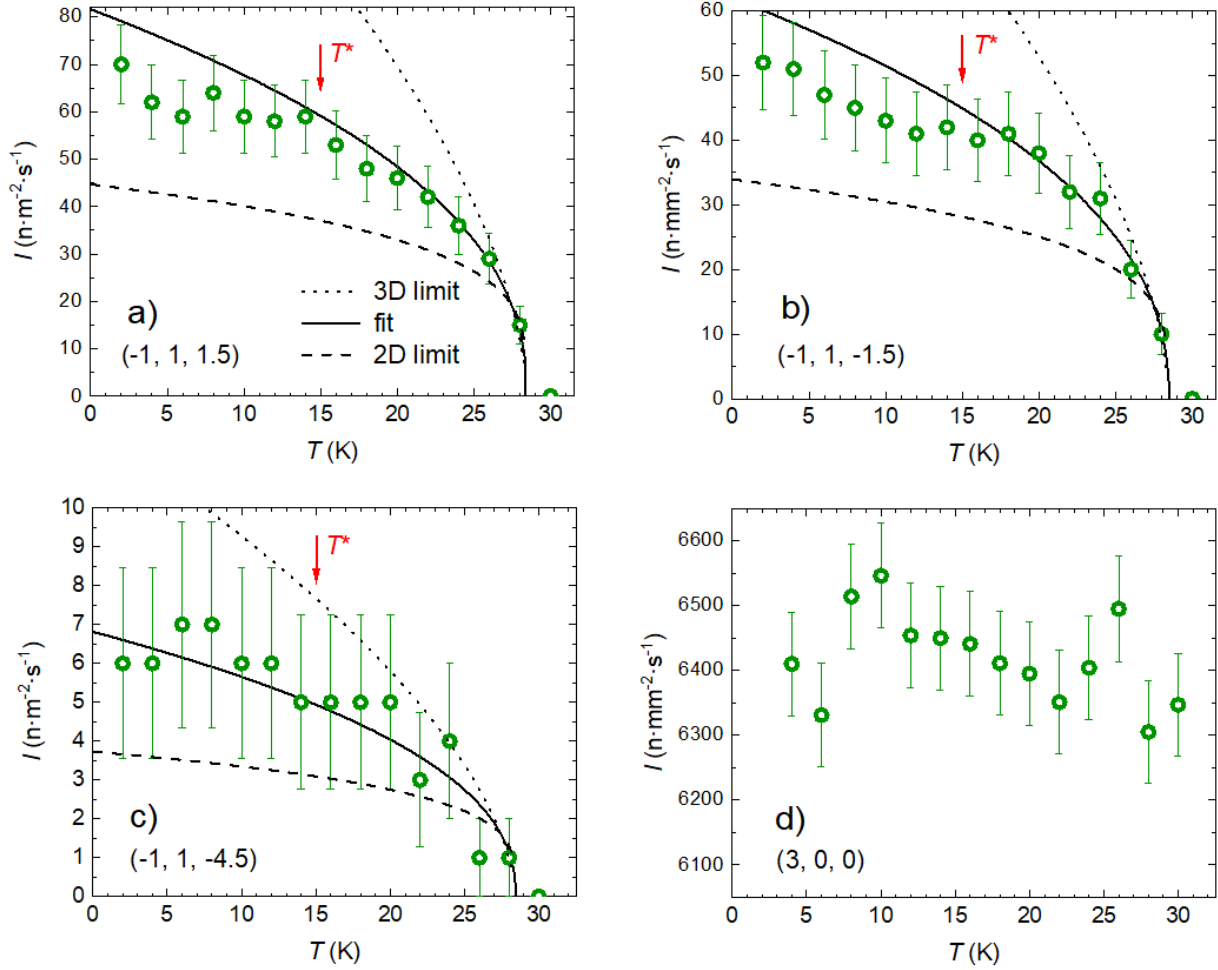

Figure 2 – Temperature evolution of intensity on magnetic reflections a) (-1, 1, 1.5), b) (-1, 1, -1.5) and c) (-1, 1, -4.5), and d) nuclear reflection (3, 0, 0). A systematic deviation of magnetic intensity from the trend below $T^* = 15$ K observed in panels a) and b) is discussed in the text. The magnetic intensity was fitted to the power-law $I \sim (T_N - T)^{2\beta}$. The fit to data below $T_N$ and its extrapolation to lower temperatures are included, together with 3D and 2D limits using the same $T_N$ value. No clear evolution of the nuclear intensity of the (3, 0, 0) reflection with temperature is followed.

The propagation vector of the magnetic structure was searched for at 2 K. First, the (0, 0, 0) propagation vector was excluded by comparing the intensities on nuclear reflections measured at 2 and 30 K; see an example of (3, 0, 0) reflection in Figure 2. No change of intensity was observed within the error bars of the measurement. Subsequently, q-scans along principal crystallographic directions of the hexagonal lattice were performed. That is, propagation vectors of a form ($k_x$, 0, 0) and (0, 0, $k_z$) were tested by mapping the reciprocal space. Surprisingly, no magnetic signal was observed in the former q-scan, contrasting expectations. DFT calculations proposed ferromagnetic zig-zag chains antiferromagnetically coupled within the basal-plane layers as the energetically most favourable solution [30]. Instead, a magnetic signal was found at positions (1, 0, -4.5) or (1, 0, -1.5).

Importantly, the propagation vector (0, 0, ½) representing the natural choice to describe the observed magnetic reflections proved to be incorrect. General diffraction conditions for the *R-3* space group dictate the observable nuclear reflections (h, k, l) to fulfill [-h+k+l=3n], where h, k, l, and n are integer numbers. Considering that the magnetic reflection (1, 0, -4.5) originates from the symmetry-allowed nuclear reflection (1, 0, -5), another magnetic reflection is expected at (1, 0, -5.5) position. That is, allowed nuclear reflection plus and minus the propagation vector should be pronounced. See the illustrative picture in Figure 3. The same would apply, e.g., for measured magnetic reflection (1, 0, -1.5) and expected reflection at (1, 0, -2.5) position. However, no intensity was observed at the respective (1, 0, -5.5) and (1, 0, -2.5) positions. Later on, such discrepancy was followed by measuring multiple other reflections, clearly excluding the propagation vector (0, 0, ½). In other words, to maintain the *R*-centering of the nuclear unit cell in the magnetic cell, the diffraction condition [-h+k+l=3n] must also be fulfilled for magnetic reflections. Therefore, considering that the propagation vector is (0, 0, ½), the reflection (1, 0, -4.5) transforms into (1, 0, -9), which does not meet the *R*-condition. Consequently, the magnetic structure must break the *R*-centering, resulting in the *P*-lattice and propagation vector $k = (1, 0, ½)$. Indeed, all subsequently measured magnetic reflections were described by this *k*.

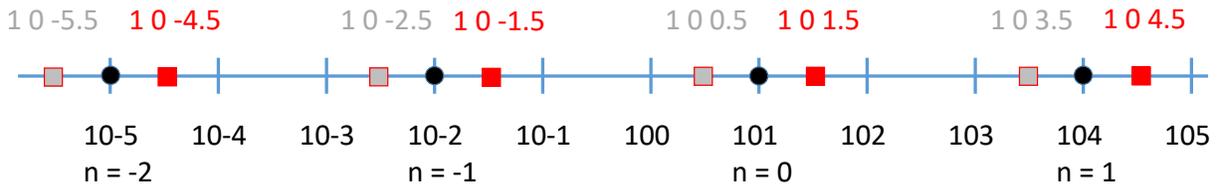

Figure 3 – Illustrative picture of selected nuclear and magnetic reflections. Black circles represent allowed nuclear reflections [-h+k+l=3n]; red squares stand for measured magnetic reflections; and grey squares indicate magnetic reflections expected for (0, 0, ½) propagation vector but not observed in the experimental data. The propagation vector (1, 0, ½) must be considered to describe the magnetic structure of VBr$_3$.

We measured 72 inequivalent magnetic reflections defined by the propagation vector *k*. A well-pronounced magnetic signal was observed at 12 reflections. Another 4 reflections showed only low magnetic intensity. The remaining 56 reflections revealed no or highly ambiguous magnetic signal. See the list of reflections in Table S1 in Supplementary Materials[46]. The magnetic origin of measured intensity was confirmed by investigating its temperature evolution at selected reflections. A smooth decrease of measured integral intensity was followed by increasing the temperature from 2 K to 30 K (Figure 2). A small residual magnetic intensity was still observed at 28 K, which is a slightly higher temperature than the magnetic ordering temperature of 26.5 K determined from the magnetization and specific heat data [30, 31]. Nevertheless, such a value is still consistent with the onset of anomalies in magnetization and specific heat, evidencing strong magnetic correlations already at slightly higher temperatures. No magnetic signal was observed at 30 K. The ordering temperature $T_N$ = 28.4 K, together with the critical exponent $\beta$ = 0.215, were estimated to fit the temperature evolution of integral intensity to the power-law $I \sim (T_N - T)^{2\beta}$. The $\beta$ value between limits for the 2D system ($\beta = 0.125$) and 3D system ($\beta = 0.3$) [47] corresponds quite well with the nature of the investigated material; that is, layered van der Waals material.

The temperature evolution of the measured integral intensity of magnetic reflections shows a systematic distinct deviation from the trend below $T^* = 15$ K deep in the AFM phase (Figure 2). We do not suppose symmetry change of the magnetic structure because the position of the magnetic reflections remained conserved and no new magnetic reflections were detected. The observed weakening of the intensities of magnetic reflections below $T^*$ can be tentatively connected with the anomalous increase of magnetic susceptibility [31], sudden increase of photoluminescence signal and emerging magnon-like signal detected by Raman spectroscopy (our unpublished data). Based on our available neutron data we cannot deduce any microscopic level conclusion about the origin of the effect.

Group theory and Isodistort software were employed to calculate the irreducible representations and related magnetic space groups of the paramagnetic *R-3* space group with $k = (1, 0, ½)$. Two irreducible representations, mLD1 (mk6t1) and mLD2LD3 (mk6t2t3), were revealed. The former representation is connected with three magnetic space groups: *P-3.1'_c[P-3]*, *P-3.1'_c[P-3]* (origin shifted by (-1/3,1/3,5/6)) and *P3.1'_c[P3]*. The three magnetic space groups lead to an antiferromagnetic alignment of magnetic moments constrained along the crystallographic c-axis. See the illustration of respective magnetic space groups in Figure S3 in Supplementary Materials. Fitting our experimental data to respective models using the FullProf suite results in a strong disagreement (Table S1 in Supplementary materials). Indeed, the alignment of magnetic moments purely along the c-axis has been excluded in the sister compounds, the orbital FM $VI_3$ [48, 49]. Based on indirect experimental evidence of orbital moment [30], a canted magnetic structure is highly likely also in $VBr_3$, considering the Kugel-Khomskii theory [50].

The second irreducible representation **mLD2LD3** allows any tilt of magnetic moments from the c-direction and contains four magnetic space groups: *P3_2.1'_c[P3_1]* (origin shifted by (1/3,1/3,0)), *P-1.1'_c[P-1]*, *P-1.1'_c[P-1]* (origin shifted by (0,0,1/2), and *P1.1'_c[P1]*. The agreement between the experimental data and most models cannot be considered better than for the mLD1 representation. Several reflections are calculated to have non-negligible magnetic intensity, which is not reproduced in the experimental data. That is, the models predict magnetic signals on reflections where zero intensity was measured; see highlighted intensities in Table S1. The lowest number of such reflections is found for magnetic space group *P-1.1'_c*.

Although this magnetic space group cannot fully describe our experimental data, it hints at the real magnetic structure. Vanadium nuclear position splits into three orbits in the magnetic space group *P-1.1'_c*. Each of $V^I$, $V^{II}$ and $V^{III}$ sites (Figure S3e) is symmetry allowed to have arbitrary magnetic moment size and direction. $V^I$ and $V^{II}$ constitute four out of six hexagonal layers in the magnetic unit cell. The $V^{III}$ site forms the remaining two layers. According to fits of magnetic moments to the model, the $V^I$ or $V^{II}$ site tends to have a zero moment. The respective layers are thus ferromagnetically ordered. $V^{III}$ moments form layers of well-separated AFM Néel chains. The magnetic unit cell of a 6c periodicity (Figure 4) is built of two AFM triple layers coupled along the c-axis. Each triple layer consists of an AFM Néel monolayer sandwiched between two AFM-coupled FM layers. Two sites, $V^I$ or $V^{II}$ and $V^{III}$ reveal almost identical moments tilted out of the *c*-direction by approximately 30 degrees.

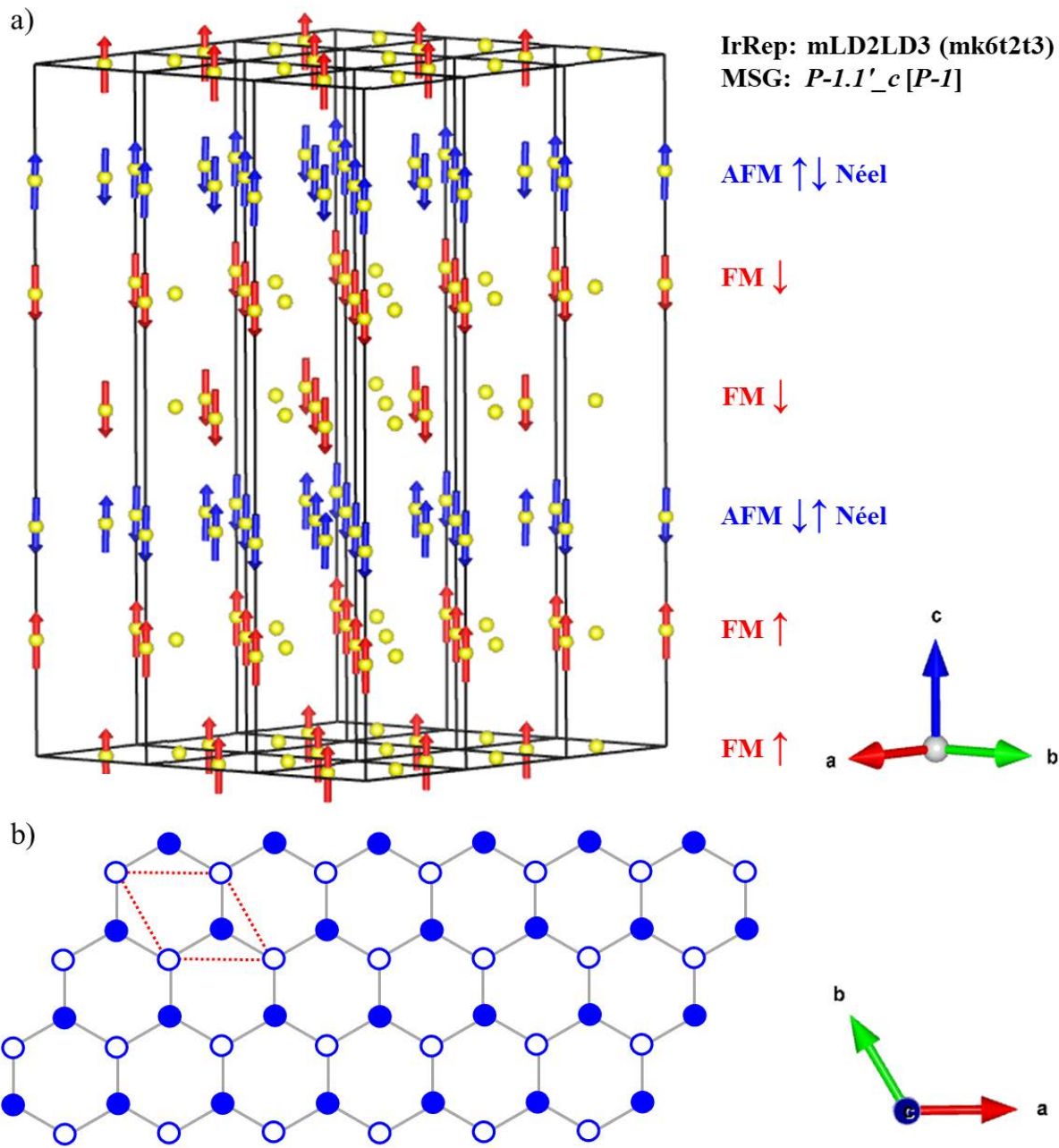

Figure 4 – The magnetic structure proposed in VBr$_3$. a) 9 magnetic unit cells are depicted to illustrate the nature of individual layers. Vanadium magnetic orbits V$^I$ (zero moments), V$^{II}$ (red arrows) and V$^{III}$ (blue arrows) are color-distinguished, creating respective AFM Néel and FM layers. b) the *c*-axis projection of a single Néel layer is provided. Red dashed lines highlight the single unit cell, while the hexagonal array better illustrates the interatomic distances and arrangement of magnetic moments. The directions of moments pointing up and down of the *ab*-plane are distinguished by full and empty circles, respectively.

Confronting this structure with previous results on VBr$_3$, a good agreement is pronounced. Our neutron diffraction study has found a very unique antiferromagnetic magnetic structure. The magnetic moments are deflected from the c-direction, similar to the sister ferromagnet VI$_3$ [48, 49]. Although the determined magnetic structure is quite different from that proposed by the theory [30], it contains predicted AFM layers. Moreover, the proposed magnetic structure could explain our high-magnetic field magnetization data, where sharp metamagnetic transitions were observed applying the field along the *c*-direction and continuous spin canting was manifested for the in-plane direction of the field [30]. Complementary studies using other microscopic techniques, such as NMR or REXS, are highly desirable in the unambiguous determination of the magnetic structure of VBr$_3$. Importantly, the magnetic structure was described in the framework of the parent paramagnetic rhombohedral *R-3* crystal structure. The previous single-crystal XRD [23] and our present neutron diffraction experiments confirmed a slight, likely monoclinic, distortion below 90 K. However, the insufficient data prevented the complete refinement of the low-temperature crystal structure. Further efforts to resolve this issue are mandatory; a synchrotron X-ray diffraction experiment is proposed to solve the low-temperature structure. Knowledge of the low-temperature crystal structure would allow us to re-analyze the magnetic neutron diffraction data and ultimately refine the magnetic structure in VBr$_3$. Also, the issue of systematic deviation of the magnetic reflection intensities from the exponential trend below $T^* = 15$ K remains unresolved and requires further experimental efforts, especially in spectroscopic experiments.

## 4. Conclusions

The magnetic structure of a VBr$_3$ was investigated using a single-crystal neutron diffraction technique. The propagation vector of the magnetic structure was determined as $k = (1, 0, ½)$ in the hexagonal notation of *R-3* crystal structure. A structural distortion below 90 K was detected by broadening the nuclear reflections; however, the low-temperature structure cannot be resolved based on the available data. Out of 72 inequivalent magnetic reflections, 16 reflections revealed countable magnetic intensity, which was confronted with model magnetic structures. The *P-1.1'_c* space group was acknowledged to likely describe the magnetic structure of VBr$_3$, despite the uncertainties related to the low-temperature crystal structure. A compensated long-period antiferromagnetic structure combining AFM Néel monolayers with ferromagnetic monolayers is proposed to explain previous experimental results. A systematic deviation of the magnetic reflection intensities from the exponential trend was detected below $T^* = 15$ K, which corresponds well with anomalies in magnetic susceptibility and magnetic excitations. The knowledge of the low-temperature crystal structure is a necessary assumption for the final drawing of the magnetic structure of the vdW AFM VBr$_3$. The complex AFM structure with propagation vector $k = (1, 0, ½)$ underlines the VBr$_3$ as a hot candidate for rich excitation spectra. The field-induced tuning of magnetic and lattice excitations to resonances, which will result in the formation of new quasiparticles, is highly expected.

## Acknowledgement

The preparation and characterisation of VBr$_3$ single crystals were carried out in MGML (http://mgml.eu/), which was supported within the Czech Research Infrastructures (project number LM2023065). The authors also acknowledge the Institute Laue-Langevin, Grenoble, France, for awarding us with an experimental time at the D10 diffractometer; experiment number 5-41-1237 (DOI: 10.5291/ILL-DATA.5-41-1237). The work was supported by the Czech Science Foundation under project number 21-06083S. We also acknowledge the Czech-French exchange programme PHC Barrande (48101TB) and the Barrande Mobility project (no. 8J24FR013).

# SUPPLEMENTARY MATERIALS to the manuscript

**Unique magnetic structure of the vdW antiferromagnet VBr$_3$**

Milan Klicpera[1,*], Ondřej Michal[1], Dávid Hovančík[1], Karel Carva[1], Oscar Ramon Fabelo Rosa[2], M. Orlita[3], Vladimír Sechovský[1], and Jiří Pospíšil[1]

[1]*Charles University, Faculty of Mathematics and Physics, Department of Condensed Matter Physics,*
*Ke Karlovu 5, 121 16 Prague 2, Czech Republic*
[2]*Institut Laue-Langevin, 71 avenue des Martyrs, CS 20156, 38042 Grenoble Cedex 9, France*
[3]*LNCMI, UPR 3228, CNRS, EMFL, Université Grenoble Alpes, 38000 Grenoble, France*

[*]Corresponding author's email address: *milan.klicpera@matfyz.cuni.cz*

ABSTRACT

VBr$_3$ is a van der Waals antiferromagnet below the Néel temperature of 26.5 K with a saturation moment of 1.2 μ$_B$/f.u. above the metamagnetic transitions detected in the in-plane and out-of-plane directions. To reveal the AFM structure of VBr$_3$ experimentally, we performed a single-crystal neutron diffraction study on a large high-quality crystal. The collected data confirmed a slight monoclinic distortion of the high-temperature rhombohedral structure below 90 K. The magnetic structure was, nevertheless, investigated within the *R-3* model. The antiferromagnetic structure propagation vector $k = (1, 0, ½)$ was revealed. In an attempt to determine the magnetic structure, 72 non-equivalent magnetic reflections were recorded. The experimental data were confronted with the magnetic space groups dictated by the *R-3* lattice symmetry and propagation vector. The best agreement between the experimental data and the magnetic structure model was obtained for the space group *P-1.1'_c*. The magnetic unit cell of the proposed unique antiferromagnetic structure with periodicity 6c is built from two identical triple layers antiferromagnetically coupled along the c axis. Each triple layer comprises a Néel antiferromagnetic monolayer sandwiched between two antiferromagnetically coupled ferromagnetic monolayers.

Keywords: VBr$_3$, van der Waals antiferromagnet, single crystal, neutron diffraction, magnetic structure, orbital moment

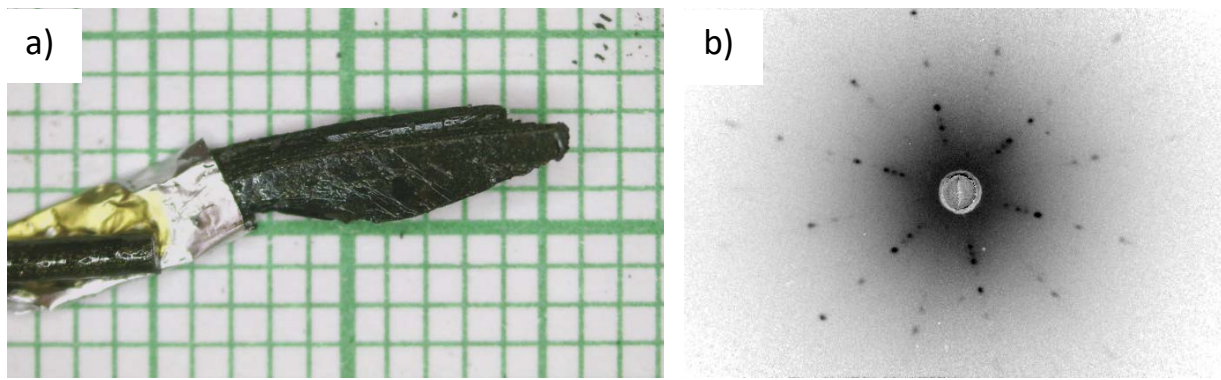

Figure S1 – VBr$_3$ single crystal used for the neutron diffraction experiment. a) The sample is enveloped tightly in aluminium foil and attached to a V holder using a GE varnish. The millimetre scale is provided for the evaluation of sample dimensions. b) The X-ray Laue pattern (negative; measured along the hexagonal axis) reveals sharp spots supporting the low mosaicity of the single crystal.

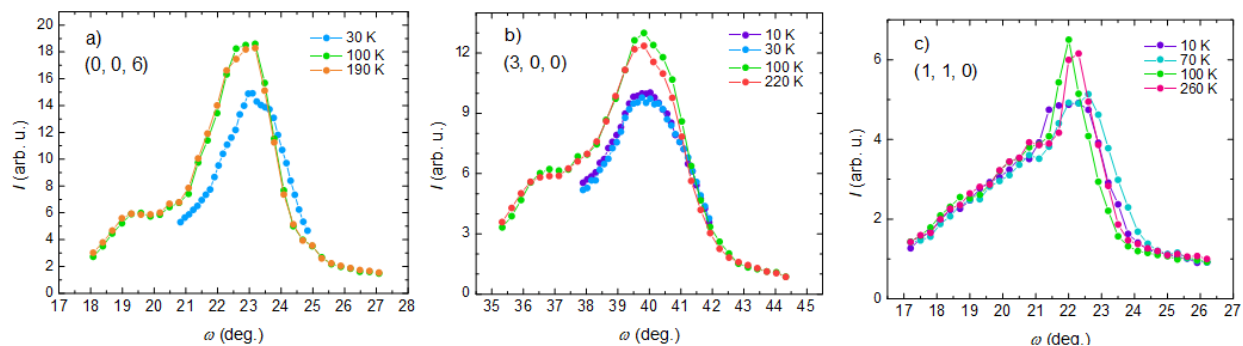

Figure S2 – Temperature evolution of $\omega$-scans around selected $R$-$3$ reflections: a) (3, 0, 0), b) (1, 1, 0) and c) (0, 0, 6). Comparing the $\omega$-scans measured above and below 90 K shows a clear difference corresponding to a structural distortion discussed in the main text. Splitting of the peaks at low temperatures is not observed due to sample mosaicity and instrumental resolution. Instead, the peaks broaden while their intensity becomes lower. The small peak at lower $\omega$-angles observed already at room temperature is ascribed to the second crystallite.

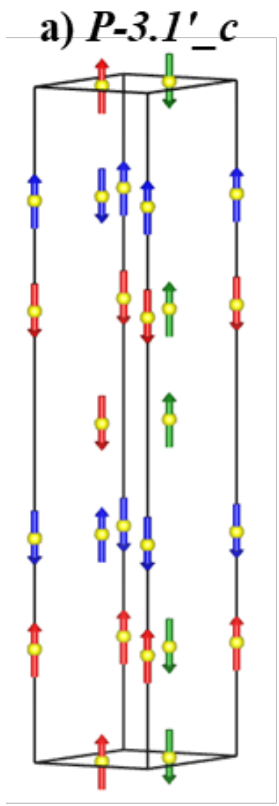 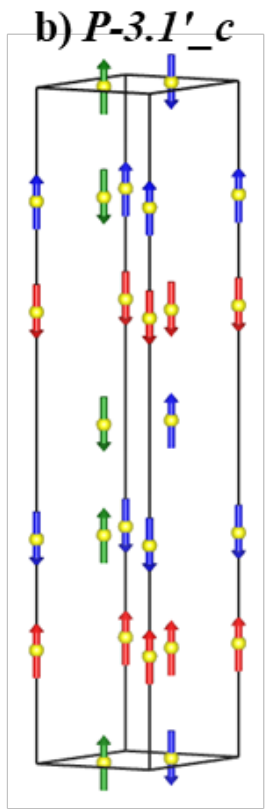 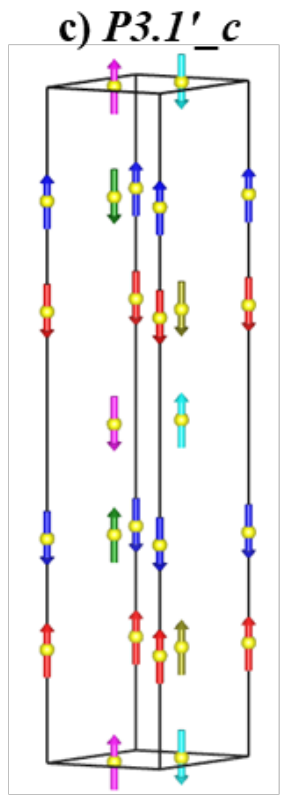 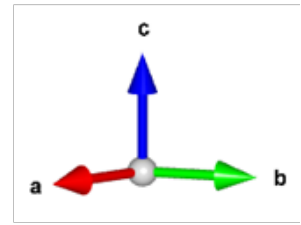
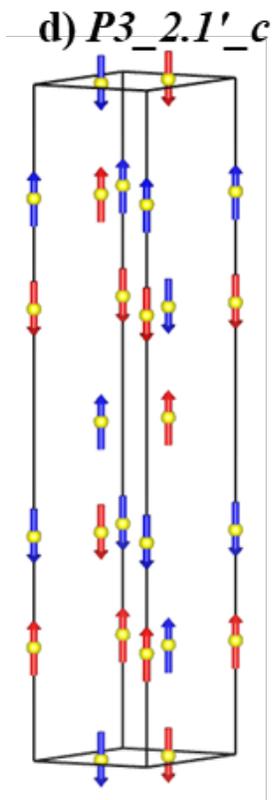 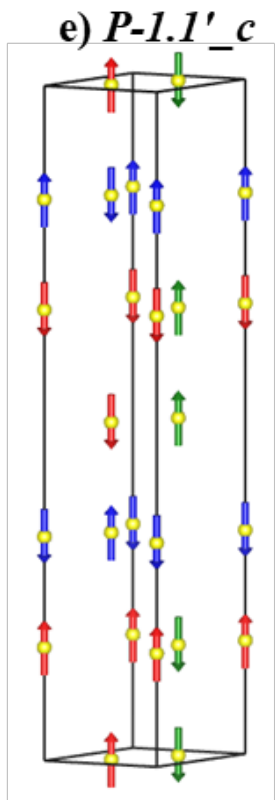 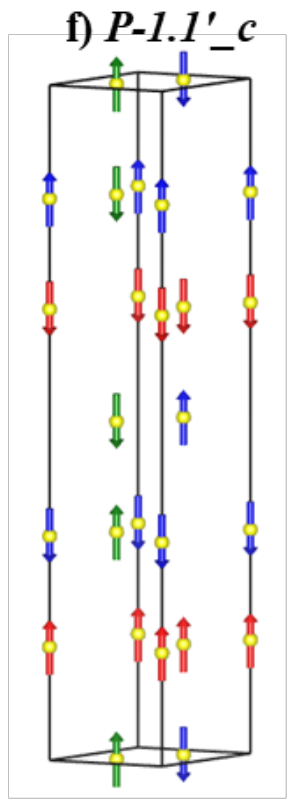 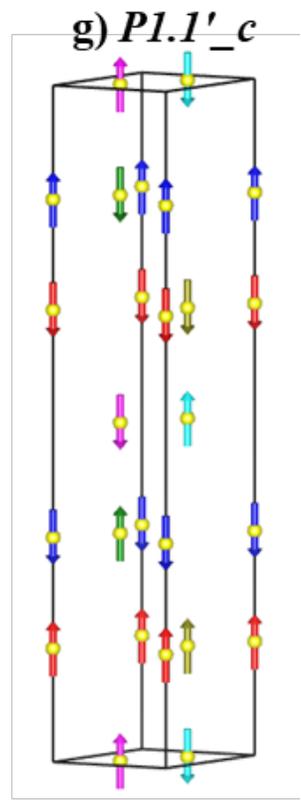

Figure S3 – Magnetic space groups calculated as subgroups of paramagnetic *R-3* with the propagation vector $k = (1, 0, ½)$. The structures a-b) and e-f) have the same symmetry but different origin: b) origin is shifted by (-1/3, 1/3, 5/6) compared to a), and f) origin is shifted by (0, 0, 1/2) compared to e). The origin of the d) structure is (1/3, 1/3, 0). The structures a-c) have magnetic moments symmetry-restricted to be parallel to the hexagonal axis. Any direction of magnetic moments is symmetry-allowed in the structures d-g); only a *c*-component is depicted for clarity and better comparison among individual structures. The paramagnetic single V site splits into three (magnetic) orbits in space groups a-b) and e-f), into two orbits in space group d), and into six orbits in space groups c) and g), indicated by colour. The moments in a single orbit are symmetry-constrained to have the same size and point along a specific direction. The mutual moment sizes and directions at different orbits are unrestrained by the time-inversion symmetry.

Table S1 – Integrated intensities measured on individual magnetic reflections $I_{exp}$ and intensities $I_{calc}$ calculated employing model magnetic structures (Figure S3). The average intensity of equivalent reflections is listed only; the list of reflections is limited up to $2θ \leq 45$ degrees. The most significant differences between $I_{calc}$ and $I_{exp}$ are highlighted in red colour. The error of $I_{exp}$ can be estimated as a square root of the value.

| h | k | l | $I_{exp}$ | $I_{calc}^{P-3.1'\_c}$ | $I_{calc}^{P3.1'\_c}$ | $I_{calc}^{P3.2'\_c}$ | $I_{calc}^{P-1.1'\_c}$ | $I_{calc}^{P1.1'\_c}$ |
|---|---|---|---|---|---|---|---|---|
| 0 | 0 | -1.5 | 0 | 0 | 0 | 0 | 0 | 0.2 |
| 0 | 0 | 1.5 | 0 | 0 | 0 | 0 | 0 | 0.2 |
| 0 | 0 | -2.5 | 3 | 18.4 | 0 | 26.3 | 4.1 | 12.9 |
| 0 | 0 | 2.5 | 3 | 18.4 | 0 | 26.3 | 4.1 | 12.9 |
| -1 | 0 | 0.5 | 0 | 0.2 | 3.4 | 0.3 | 0 | 1.4 |
| 0 | 1 | -0.5 | 1 | 4.1 | 11.0 | 0.3 | 7.9 | 12.9 |
| 1 | 0 | -0.5 | 0 | 0.2 | 3.4 | 0.3 | 0 | 1.4 |
| 1 | 0 | 0.5 | 0 | 4.1 | 11.2 | 0.3 | 12.2 | 13.2 |
| 0 | 0 | 3.5 | 0 | 15.1 | 0 | 20.6 | 3.4 | 8.6 |
| 0 | 0 | -3.5 | 0 | 15.1 | 0 | 20.6 | 3.4 | 8.6 |
| -1 | 0 | 1.5 | 45 | 10.0 | 38.0 | 12.7 | 47.0 | 33.8 |
| -1 | 1 | 1.5 | 59 | 10.3 | 38.8 | 13.5 | 35.4 | 53.9 |
| -1 | 1 | -1.5 | 46 | 9.9 | 36.3 | 12.5 | 43.3 | 43.0 |
| 0 | 1 | -1.5 | 44 | 10.1 | 35.6 | 13.1 | 42.8 | 37.3 |
| 1 | 0 | -2.5 | 0 | 4.9 | 7.8 | 0.4 | 6.9 | 8.9 |
| -1 | 0 | 2.5 | 0 | 4.9 | 7.8 | 0.4 | 6.9 | 8.9 |
| -1 | 1 | 2.5 | 0 | 0.1 | 1.8 | 0.2 | 0.1 | 1.7 |
| 0 | 1 | -2.5 | 0 | 0.1 | 1.8 | 0.2 | 0.1 | 4.1 |
| 0 | 0 | 4.5 | 0 | 0 | 0 | 0 | 0 | 0.2 |
| 0 | 0 | -4.5 | 0 | 0 | 0 | 0 | 0 | 0.2 |
| -1 | 0 | -3.5 | 0 | 4.7 | 3.5 | 0.4 | 4.8 | 7.0 |
| -1 | 0 | 3.5 | 0 | 0.1 | 1.3 | 0.1 | 0.1 | 2.0 |
| 1 | 0 | 3.5 | 0 | 4.7 | 3.5 | 0.4 | 4.8 | 7.0 |
| -1 | 1 | -3.5 | 0 | 0.1 | 1.3 | 0.1 | 0.1 | 3.4 |

Table S1 – *Continuation* – Integrated intensities measured on individual magnetic reflections $I_{exp}$ and intensities $I_{calc}$ calculated employing model magnetic structures (Figure S3). The average intensity of equivalent reflections is listed only; the list of reflections is limited up to $2\theta \leq 45$ degrees. The most significant differences between $I_{calc}$ and $I_{exp}$ are highlighted in red colour. The error of $I_{exp}$ can be estimated as a square root of the value.

| h | k | l | $I_{exp}$ | $I_{calc}^{P\text{-}3.1'\_c}$ | $I_{calc}^{P3.1'\_c}$ | $I_{calc}^{P3.2'\_c}$ | $I_{calc}^{P\text{-}1.1'\_c}$ | $I_{calc}^{P1.1'\_c}$ |
|---|---|---|---|---|---|---|---|---|
| 0 | 0 | 5.5 | 0 | 11.7 | 0 | 17.1 | 2.7 | 8.9 |
| 0 | 0 | -5.5 | 0 | 11.7 | 0 | 17.1 | 2.7 | 8.9 |
| -1 | 0 | 4.5 | 12 | 9.4 | 11.3 | 11.5 | 24.8 | 21.5 |
| 1 | 0 | -4.5 | 12 | 9.4 | 11.3 | 11.5 | 24.8 | 21.5 |
| -1 | 0 | -4.5 | 13 | 10.3 | 11.3 | 13.9 | 6.5 | 17.5 |
| 1 | 0 | 4.5 | 9 | 10.3 | 11.3 | 13.9 | 6.5 | 17.5 |
| -2 | 1 | 0.5 | 0 | 5.2 | 7.6 | 7.3 | 3.1 | 5.6 |
| 1 | 1 | -0.5 | 0 | 5.1 | 7.6 | 7.2 | 3.9 | 3.9 |
| 1 | 1 | -1.5 | 0 | 0.2 | 0 | 0.1 | 0 | 0.3 |
| 1 | 1 | 1.5 | 0 | 0.2 | 0 | 0.1 | 0 | 0.4 |
| -1 | 2 | 1.5 | 0 | 0.2 | 0 | 0.1 | 0 | 0.4 |